\documentclass[11pt]{article}
\usepackage{amsmath}
\usepackage{amssymb}
\usepackage{bbm}
\usepackage{epsfig}
\allowdisplaybreaks[4]
\begin{document}

\title{Strong flavour changing effective operator contributions to single top 
quark production}
\author{P.M. Ferreira~\footnote{ferreira@cii.fc.ul.pt},
R. Santos~\footnote{rsantos@cii.fc.ul.pt}\\ 
Centro de F\'{\i}sica Te\'orica e Computacional, Faculdade de Ci\^encias,\\
Universidade de Lisboa, Av. Prof. Gama Pinto, 2, 1649-003 Lisboa, Portugal }
\date{January, 2006} 
\maketitle
\noindent
{\bf Abstract.} We study the effects of dimension six effective operators on the
production of single top quarks at the LHC. The operator set considered includes
terms with effective gluon interactions and four-fermion terms. Analytic
expressions for the several partonic cross sections of single top production 
will be presented, as well as the results of their integration on the parton
density functions. 
\vspace{1cm}

\section{Introduction}

The top quark~\cite{rev} is the heaviest particle thus far discovered. Its large
mass makes it a natural laboratory to investigate deviations from Standard Model
(SM) physics. Recently~\cite{nos} we undertook a model-independent study of 
possible new physics effects on the phenomenology of the top quark. To this 
effect we considered a set of dimension six effective operators and analyzed its
impact on observable quantities related to the top quark, such as its width or
the cross section for single top quark production at the LHC. This procedure -
the use of effective operators of dimension larger than four, the complete list
of dimension five and six operators obtained in reference~\cite{buch} - has been
widely used to study the top particle. In refs.~\cite{whis} the contributions 
from several dimension five and six operators for top quark physics, both at the
Tevatron and the LHC, were studied. The $W\,t\,b$ vertex was studied in great 
detail by the authors of ref.~\cite{saav}. Because many proposals for theories
that extend the SM (such as two Higgs doublet models or Supersymmetry) have 
potentially large contributions to flavour changing neutral currents, they have
been the subject of many detailed studies, such as those found in~\cite{fcnc}. 
Recent examples concerning single top production in supersymmetric models may
be found in~\cite{sola}. NLO and threshold corrections to flavour changing 
effective operators involving the top quark were studied in~\cite{liu}. A 
different type of study, using four-fermion operators to tackle the issue of 
$t\bar{t}$ production, was undertaken in~\cite{4f}.

In ref.~\cite{nos} we chose a particular set of dimension six operators and
studied its effects on the properties of the top quark. Our choice was 
motivated by the following arguments: it included several operators already 
studied by other authors, albeit not in conjunction; those operators model 
possible effects arising from several different interesting extensions of the 
SM; they had little or no impact on phenomena occurring at energy 
scales inferior to the LHC's; and finally, the operators chosen involved flavour
changing strong interactions with a single top quark and a gluon. Our philosophy
in~\cite{nos} was also somewhat different from that of most previous works in 
this field, in that we presented, whenever possible, analytical expressions. Our
aim was, and is, to provide our experimental colleagues with formulae they can 
use directly in their Monte Carlo simulations. 

In~\cite{nos} we studied the simplest physical consequences of the operator set
we chose: its contributions to the top's width; the possibility of direct top 
production at the LHC; their effects on the production of a single top at the
LHC via interference with the SM processes. In this work we will apply these
same operators to more complicated processes of single top production, namely
via partonic channels such as $g\,g\,\rightarrow\,t\,\bar{c}$, $g\,u\,
\rightarrow\,g\,t$, $q\,\bar{q}\,\rightarrow\,t\,\bar{u}$ and others. We will 
also expand our operator set, considering three different types of four-fermion 
operators, which have great relevance for eight different processes of single 
top production. Their interference with the gluonic operators will also be 
studied.

This paper is structured as follows: in section~\ref{sec:eff} we will review the
effective operator formalism, the criteria behind our choice of operators in 
ref.~\cite{nos} and the results therein obtained. We will also introduce the 
new four-fermion operators and explain the logic behind their choice, and how 
their presence is demanded by the very equations of motion. In 
section~\ref{sec:res} we present the calculations of the partonic cross sections
for the processes $g\,g\,\rightarrow\,t\,\bar{u}$ and $g\,u\,\rightarrow\,g\,t$,
as well as, after integration in the partonic density functions (pdf's), their
expected values at the LHC. In section~\ref{sec:4f} we will compute partonic
cross sections of the form $q\,q^\prime\,\rightarrow\,t\,q^{\prime\prime}$, 
where the four-fermion operators are now of crucial importance. Finally, in 
section~\ref{sec:conc} we will make a general discussion of the results obtained
and draw some conclusions. 
 
\section{Effective operators and the top quark}
\label{sec:eff}

The effective operator approach is based on the assumption that, at a given 
energy scale $\Lambda$, physics effects beyond those predicted by the SM make 
themselves manifest. We describe this by assuming the lagrangean
\begin{equation}
{\cal L} \;\;=\;\; {\cal L}^{SM} \;+\; \frac{1}{\Lambda}\,{\cal L}^{(5)} \;+\;
\frac{1}{\Lambda^2}\,{\cal L}^{(6)} \;+\; O\,\left(\frac{1}{\Lambda^3}\right)
\;\;\; ,
\label{eq:l}
\end{equation}
where ${\cal L}^{SM}$ is the SM lagrangean and ${\cal L}^{(5)}$ and 
${\cal L}^{(6)}$ are all of the dimension 5 and 6 operators which, like 
${\cal L}^{SM}$, are invariant under the gauge symmetries of the SM. The 
${\cal L}^{(5)}$ terms break baryon and lepton number conservation, and are thus
not usually considered. This leaves us with the ${\cal L}^{(6)}$ operators, some
of which, after spontaneous symmetry breaking, generate dimension five terms.
The list of dimension six operators is quite vast~\cite{buch}, therefore some 
sensible criteria of selection are needed. Underlying all our work is the desire
to study a new possible type of physics, flavour changing strong interactions. 
The first criterion is to choose those ${\cal L}^{(6)}$ operators that have no 
sizeable impact on low energy physics (below the TeV scale, say). Another 
criterion was to only consider operators with a single top quark, since  
we will limit our studies to processes of single top production. Finally, we 
will restrict ourselves to operators with gluons, or four-fermion ones. No 
effective operators with electroweak gauge bosons will be considered. 

\begin{figure}[ht]
\epsfysize=12cm
\centerline{\epsfbox{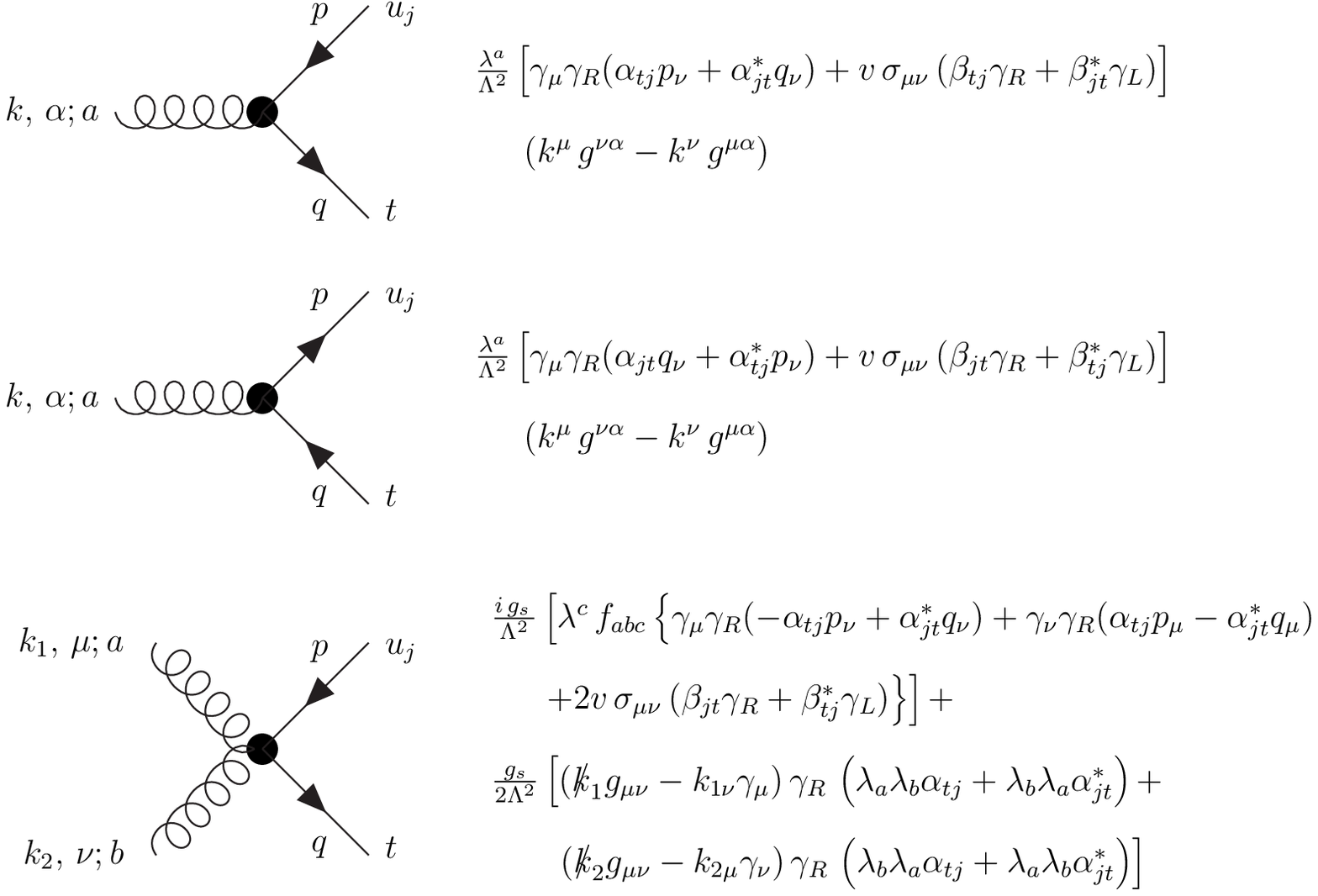}}
\caption{Feynman rules for anomalous gluon vertices.}
\label{fig:feynrul}
\end{figure}

The gluon operators that survive these criteria are but two, which, in the 
notation of ref.~\cite{buch}, are written as
\begin{align}
{\cal O}_{uG} &= \;\;i\,\frac{\alpha_{ij}}{\Lambda^2}\,\left(\bar{u}^i_R\,
\lambda^a\, \gamma^\mu\,D^\nu\,u^j_R\right)\,G^a_{\mu\nu} \nonumber
\vspace{0.2cm} \\
{\cal O}_{uG\phi} &= \;\;\frac{\beta_{ij}}{\Lambda^2}\,\left(\bar{q}^i_L\,
\lambda^a\, \sigma^{\mu\nu}\,u^j_R\right)\,\tilde{\phi}\,G^a_{\mu\nu} \;\;\; .
\label{eq:op}
\end{align}   
$q_L$ and $u_R$ are spinors (a left quark doublet and up-quark right singlet of
$SU(2)$, respectively), $\tilde{\phi}$ is the charge conjugate of the Higgs 
doublet and $G^a_{\mu\nu}$ is the gluon tensor. $\alpha_{ij}$ and $\beta_{ij}$
are complex dimensionless couplings, the $(i,j)$ being flavour indices. 
According to our criteria, one of these indices must belong to the third 
generation. After spontaneous symmetry breaking the neutral component of the 
field $\phi$ acquires a vev ($\phi_0\,\rightarrow\,\phi_0\,+\,v$, with $v\,=\,
246/\sqrt{2}$ GeV) and the second of these operators generates a dimension 
five term. The lagrangean for new physics thus becomes
\begin{align}
{\cal L}\;\; =&\;\;\; \alpha_{tu}\,{\cal O}_{tu}\;+\; \alpha_{ut}\,{\cal O}_{ut}
\;+\; \beta_{tu}\,{\cal O}_{tu\phi}\;+\;\beta_{ut}\,{\cal O}_{ut\phi}\;+\;
\mbox{h.c.} \nonumber \vspace{0.2cm} \\
 =& \;\;\;\frac{i}{\Lambda^2}\,\left[\alpha_{tu}\,\left(\bar{t}_R\,\lambda^a\,
\gamma^\mu \,D^\nu\,u_R\right)\;+\;  \alpha_{ut}\,\left(\bar{u}_R\,\lambda^a\,
\gamma^\mu\, D^\nu\,t_R\right)\right]\,G^a_{\mu\nu} \;\;\;+ \nonumber
\vspace{0.2cm} \\
 & \;\;\;\frac{v}{\Lambda^2}\,\left[\beta_{tu}\,\left(\bar{t}_L\,\lambda^a\,
\sigma^{\mu\nu}\,u_R\right)\;+\; \beta_{ut}\,\left(\bar{u}_L\,\lambda^a\,
\sigma^{\mu\nu}\,t_R\right)\right]\,G^a_{\mu\nu} \;\;+\;\; \mbox{h.c.}
\;\;\;.
\label{eq:lf}
\end{align}
This lagrangean describes new vertices of the form $g\,\bar{t}\,u$ ($g\,t\,
\bar{u}$) and $g\,g\,\bar{t}\,u$ ($g\,g\,t\,\bar{u}$). We will also consider an 
analogous lagrangean (with new couplings $\alpha_{tc}$, $\beta_{ct}$, \ldots) 
for vertices of the form $g\,\bar{t}\,c$ ($g\,t\,\bar{c}$) and $g\,g\,\bar{t}\,
c$ ($g\,g\,t\,\bar{c}$). Notice how the operators with $\beta$ couplings 
correspond to a chromomagnetic momentum for the $t$ quark. Several extensions of
the SM, such as supersymmetry and two Higgs doublet models, may generate 
contributions to this type of operator~\cite{chro}. The Feynman rules for
these anomalous vertices are shown in figure~\eqref{fig:feynrul}, with quark 
momenta following the arrows and incoming gluon momenta. The double gluon vertex
was not considered in ref.~\cite{nos} because it was not necessary there but, as
we shall shortly see, it is of vital importance for this paper. 

In ref.~\cite{nos} we calculated the effect of these operators on the width of
the quark top. They allow for the decay $t\,\rightarrow\,u\,g$ ($t\,\rightarrow
\,c\,g$) (which is also possible in the SM, albeit at higher orders), and the 
corresponding width is given by
\begin{align}
\Gamma (t \rightarrow u g) &=\;  \frac{m^3_t}{12 \pi\Lambda^4}\,\Bigg\{ m^2_t
\,\left|\alpha_{ut}  +  \alpha^*_{tu}\right|^2 \,+\, 16 \,v^2\, \left(\left|
\beta_{tu} \right|^2 + \left| \beta_{ut} \right|^2 \right) \;\;\; +
\vspace{0.3cm} \nonumber \\
 & \hspace{2.2cm}\, 8\, v\, m_t\,\mbox{Im}\left[ (\alpha_{ut}  + \alpha^*_{tu})
\, \beta_{tu} \right] \Bigg\}
\label{eq:wid}
\end{align}
and an analogous expression for $\Gamma (t \rightarrow c g)$. In this 
expression, and throughout the entire paper, we will consider all quark masses, except the top's, equal to zero; the imprecision introduced by this 
approximation is extremely small, as we verified having performed the full 
calculations. Direct top production is also possible with these new vertices 
(meaning, the production of a top quark from partonic reactions such as $g\,u\,
\rightarrow\,t$ or $g\,c\, \rightarrow\,t$), and the corresponding cross section
at the LHC is given by
\begin{equation}
\sigma(p\,p\,\rightarrow\,t)\;\;=\;\;\sum_{q\,=\,u,c}\,\Gamma (t\,\rightarrow\,q
\,g)\,
\frac{\pi^2}{m_t^2}\;\int^1_{m^2_t/E_{CM}^2}\frac{2\, m_t}{E_{CM}^2\, x_1} f_g
(x_1)\, f_q (m^2_t/(E_{CM}^2\, x_1)) \, dx_1 \;\;\; .
\label{eq:ppt}
\end{equation}
In this expression $E_{CM}$ is the proton-proton center-of-mass energy (14 TeV 
at the LHC) and $f_g$ and $f_q$ are the parton density functions of the gluon
and quark, respectively. It is not surprising that this cross section is 
proportional to the partial widths $\Gamma (t\,\rightarrow\,q\,g)$ - after all,
the amplitudes for the decay $t\,\rightarrow\,u\,g$ or for direct top production
via the channel $u\,g\,\rightarrow\,t$ are closely related by a time inversion
transformation. Numerical results for these quantities were obtained in 
ref.~\cite{nos}, where we also derived bounds on the values of the $\{\alpha\,,
\,\beta\}$ couplings.   

Notice how both the top width~\eqref{eq:wid} and the cross 
section~\eqref{eq:ppt} depend on $\Lambda^{-4}$. There are processes with a 
$\Lambda^{-2}$ dependence, namely the interference terms between the anomalous
operators and the SM diagrams of single top quark production, via the exchange
of a W gauge boson - processes like $u\,\bar{d}\,\rightarrow\,t\,\bar{d}$. They
were studied in ref.~\cite{nos} in detail, and we discovered that, due to a 
strong CKM suppression, the contributions from the anomalous vertices are 
extremely small. We will come back to this point later. 

Now, the operators that compose the lagrangean~\eqref{eq:lf} are not, in fact,
completely independent. If one performs integrations by parts and uses the
fermionic equations of motion~\cite{buch,grz}, one obtains the following 
relations between them:
\begin{align}
{\cal O}^{\dagger}_{ut} &= {\cal O}_{tu}\;-\;\frac{i}{2} (\Gamma^{\dagger}_u\,
{\cal O}^{\dagger}_{u t \phi} \,+\, \Gamma_u \,{\cal O}_{t u \phi}) \nonumber \\
{\cal O}^{\dagger}_{ut} &= {\cal O}_{tu}\;-\;i\, g_s\, \bar{t}\, \gamma_{\mu}\,
\gamma_R\, \lambda^a\,u\, \sum_i  (\bar{u}^i\, \gamma^{\mu}\, \gamma_R\,
\lambda_a u^i\,+\, \bar{d}^i\, \gamma^{\mu}\, \gamma_R\, \lambda_a\, d^i)
\;\;\; ,
\label{eq:rel}
\end{align}
where $\Gamma_u$ are the Yukawa couplings of the up quark and $g_s$ the strong 
coupling constant. In the second of these equations we see the appearance of
four-fermion terms, indicating that they have to be taken into account in these 
studies. Equations~\eqref{eq:rel} then tell us that there are two relations
between the several operators, which means that we are allowed to set two of the
couplings to zero. 

A careful analysis of the operators listed in~\cite{buch} leads us to consider 
three types of four-fermion operators:

\begin{itemize} 
\item{Type 1, 
\begin{equation}
{\cal O}_{u_1}\;\;=\;\; \frac{g_s\,\gamma_{u_1}}{\Lambda^2}
\left(\bar{t}\, \lambda^a\,\gamma^{\mu}\, \gamma_R\, u\right)\,\left(\bar{q}
\, \lambda^a\,\gamma_{\mu}\, \gamma_R\, q\right)\;+\;\mbox{h.c.} \;\;\; ,
\end{equation}
where $q$ is any given quark, other than the top;}
\item{Type 2, 
\begin{equation}
{\cal O}_{u_2}\;\;=\;\; \frac{g_s\,\gamma_{u_2}}{\Lambda^2}
\left[\left(\bar{t}\, \lambda^a\, \gamma_L\, u^\prime\right)\,\left(
\bar{u}^{\prime\prime}\,\lambda^a\, \gamma_R\, u\right) \; + \; \left(\bar{t}\, 
\lambda^a\, \gamma_L\, d^\prime\right)\,\left(\bar{d}^{\prime\prime}\,\lambda^a
\,\gamma_R\, u\right) \right] \;+\;\mbox{h.c.} \;\;\; ,
\end{equation}
with down and up quarks from several possible generations, excluding the top 
once more;}
\item{Type 3,
\begin{equation}
{\cal O}_{u_3}\;\;=\;\; x\,\frac{g_s\,\gamma_{u_3}}{\Lambda^2}
\left[\left(\bar{t}\, \lambda^a\, \gamma_R\, u\right)\,\left(
\bar{b}\,\lambda^a\, \gamma_R\, d^\prime\right) \; - \; \left(\bar{t}\,
\lambda^a\, \gamma_R\, d^\prime\right)\,\left(\bar{b}\,\lambda^a\,\gamma_R\,u
\right) \right] \;+\;\mbox{h.c.} \;\;\; ,
\label{eq:ga31}
\end{equation}
and also,
\begin{equation}
\frac{g_s\,\gamma_{u_3}^*}{\Lambda^2}
\left[\left(\bar{t}\, \lambda^a\, \gamma_L\, u\right)\,\left(
\bar{d}^\prime\,\lambda^a\, \gamma_L\, d^{\prime\prime}\right) \; - \; \left(
\bar{t}\, \lambda^a\, \gamma_L\, d\right)\,\left(\bar{d}^\prime\,\lambda^a\,
\gamma_L\,u^{\prime\prime}\right) \right] \;+\;\mbox{h.c.} \;\;\; .
\label{eq:ga32}
\end{equation}
}
\end{itemize}
The $\gamma_u$'s are complex couplings. We of course consider identical 
operators for the case of flavour changing interactions with the $c$ quark. In 
the notation of ref.~\cite{buch} these operators correspond, respectively, to 
$\bar{R}R\bar{R}R$, $\bar{L}R\bar{R}L$ and $\bar{L}R\tilde{(\bar{L}R)}$, in the 
octet configuration. We could have also considered the singlet operators but, 
since their spinorial structure is identical to these (lacking only the 
Gell-Mann matrices) we opted to leave them out. The presence of the $\lambda^a$ 
in these operators also signals their origin within the strong interaction 
sector, in line with our aim of studying strong flavour changing effects. For 
this reason, and for an easier comparison between the effects of the several 
operators, we included, in the definitions of the four-fermion terms above, an
overall factor of $g_s$. The relative signs and disposition of quark 
spinors in the operators of types 2 and 3 are a reflex of their particular 
structure, emerging as they do from combinations of $SU(2)$ singlets and 
doublets. In eq.~\eqref{eq:ga31} we included a multiplicative factor of ``$x$";
as can be seen from that equation, the four-fermion operator in question
concerns {\em only} the bottom quark - unlike the operator in 
eq.~\eqref{eq:ga32}, for which the flavour of the down-type quarks was left 
free. This means that for processes involving the bottom quark there will be 
more contributions to the amplitude than for non-bottom quarks. We will reflect
this in our results by expressing them in terms of $x$ - the reader will then 
know that if that particular cross section contribution involves a bottom quark
one must set $x\,=\,1$, if not, then $x\,=\,0$. 

We emphasize that many possible operators with a single top quark were 
left out, due to another of our criteria, that low energy physics be not
affected. For instance, we could have considered a type 2 operator of the form 
$(\bar{Q}_L\,\lambda^a\,u_R)\,(\bar{u}_R\,\lambda^a\,Q^\prime_L)$, where $Q_L$ 
is the quark doublet of the third generation, $Q_L\,=\,(t_L\,,\,b_L)$ and 
$Q^\prime_L$ is a quark doublet of another generation. This would produce two 
terms in the lagrangean, namely
\begin{equation}
\left(\bar{t}\, \lambda^a\, \gamma_R\, u\right)\,\left( \bar{u}\,\lambda^a\, 
\gamma_L\, u^\prime\right) \; + \; \left(\bar{b}\, \lambda^a\, \gamma_R\, 
u\right)\,\left(\bar{u}\,\lambda^a \,\gamma_R\, d^\prime\right)\;\;\; ,
\end{equation}
and the second term in this expression has no bearing on top physics. It would
only impact bottom physics, for instance, and thus its effects are already 
immensely constrained by the existing data. Finally, the reader will notice that
we considered the same constants $\gamma_1$, $\gamma_2$ and $\gamma_3$ 
regardless of the flavour structure of the four-fermion operators. Having 
distinguished earlier on between ${\cal O}_{ct}$ and ${\cal O}_{tc}$, for 
instance, with the couplings $\alpha_{ct}$ and $\alpha_{tc}$, we should do the 
same here for consistency. However, that would introduce an enormous number of
unconstrained parameters in the calculations, which would constitute a needless
complication. We chose this simpler approach.

\section{Cross sections for $g\,g\,\rightarrow\,t\,\bar{u}$ and $g\,u\,
\rightarrow\,g\,t$}
\label{sec:res}

\begin{figure}[ht]
\epsfysize=7cm
\centerline{\epsfbox{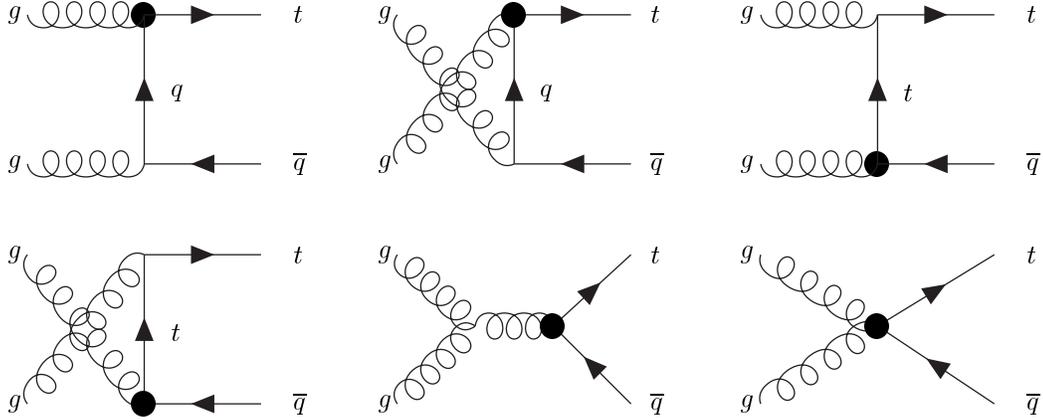}}
\caption{Feynman diagrams for the two-gluon channel.}
\label{fig:gg}
\end{figure}

In ref.~\cite{nos} we considered the contributions from the anomalous gluon 
operators to the simpler processes of single top production. We now present the
results for more elaborate reactions, such as $g\,g\,\rightarrow\,t\,\bar{u}$. 
There are six Feynman diagrams contributing to this partonic cross section, 
shown in fig.~\eqref{fig:gg}. With the Feynman rules shown in 
fig.~\eqref{fig:feynrul} it is a simple, if laborious, task to calculate the 
cross section $\sigma (g\,g\,\rightarrow\,t\,\bar{u})$. However, one important
calculational detail warrants a special mention, given how seldom it is 
mentioned in the literature. For any process involving a single gluon with 
polarization $\epsilon^{a,i}_\mu (k)$, the calculation of the squared amplitude 
will involve the sum of the gluon polarizations, $\sum_i\,\epsilon^{a,i}_\mu (k)
{\epsilon^*}^{b,i}_\nu (k)$. The use of the ``Feynman trick" from QED is allowed
and one may replace this sum by $-\,\delta^{ab}\,g_{\mu\nu}$. However, for any 
process involving two or more gluons, this is no longer possible. Instead, one
must use the more complex expression~\cite{glsu}
\begin{equation}
\sum_{spins}\,\epsilon^{a}_\mu \,{\epsilon^*}^{a}_\nu  \;\;\;=\;\;\; \delta^{ab}
\,\left[-\,g_{\mu\nu}\;+\;\frac{2}{s}\left({p_1}_\mu\,{p_2}_\nu\,+\,{p_1}_\nu\,
{p_2}_\mu\right)\right] \;\;\;\ ,
\label{eq:gls}
\end{equation}
where $p_1$ and $p_2$ are the 4-momenta of the incoming particles (even if they
are not gluons). The reason for this more complicated structure is the 
non-abelian nature of the theory. The extra terms in eq.~\eqref{eq:gls} arise
from the need to introduce Fadeev-Popov ghosts in the quantification of QCD. 
Failure to use the full structure of eq.~\eqref{eq:gls} will result on a 
break of unitarity and negative cross sections. 

For this process, then, the full calculation yields
\begin{equation}
\frac{d\,\sigma(g\,g\rightarrow t\,\bar{u})}{dt}\;\; =\;\; -\,
\frac{g_s^2}{4\,m_t^3}\; \frac{F_{gg}}{u\,t\,s^3\,(s + t)^2\,(s + u)^2}
\; \Gamma (t\,\rightarrow\,u\,g) \;\;\; ,
\label{eq:gg}
\end{equation}
where
\begin{align}
F_{gg} \;=&\;\; 4\,s^2\,t\,{\left( s + t \right) }^3 \left( s^2 + 2\,s\,t + 
2\,t^2 \right)  + s\,{\left( s + t \right) }^2 \left( 4\,s^4 + 11\,s^3\,t + 
48\,s^2\,t^2 + 52\,s\,t^3 + 18\,t^4 \right) u \nonumber \vspace{0.3cm} \\
 & + 2\,\left( s + t \right) \, \left( 10\,s^5 + 27\,s^4\,t + 69\,s^3\,t^2 + 
90\,s^2\,t^3 + 45\,s\,t^4 + 9\,t^5 \right) \,u^2 \nonumber \vspace{0.3cm} \\
 & + \left( s + t \right) \,\left( 44\,s^4 + 115\,s^3\,t + 203\,s^2\,t^2 + 
162\,s\,t^3 + 36\,t^4 \right) \,u^3 \nonumber \vspace{0.3cm} \\
 & + 2\,\left( 26\,s^4 + 85\,s^3\,t + 135\,s^2\,t^2 + 99\,s\,t^3 + 27\,t^4 
\right) \,u^4 + 4\,\left( 2\,s + t \right) \, \left( 4\,s^2 + 9\,s\,t + 9\,t^2 
\right) \,u^5 \nonumber \vspace{0.3cm} \\
 & + 2\,\left( 4\,s^2 + 9\,s\,t + 9\,t^2 \right) \,u^6 \;\;\; .
\end{align}
Remarkably, as in the case of the cross section for direct top production, this
result is proportional to the width $\Gamma (t\,\rightarrow\,u\,g)$. 

\begin{figure}[ht]
\epsfysize=7cm
\centerline{\epsfbox{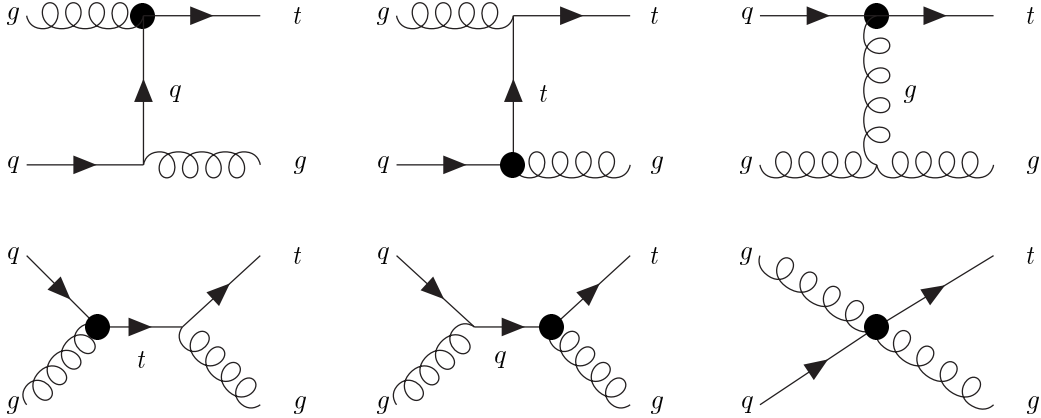}}
\caption{Feynman diagrams for the gluon-quark channel.}
\label{fig:gq}
\end{figure}

For the process $g\,u\,\rightarrow\,g\,t$ the procedure is very similar. We also
have six diagrams, shown in fig.~\eqref{fig:gq}. The cross section for this  
process is clearly related to that of the previous one by crossing. We obtain
\begin{equation}
\frac{d\,\sigma(g\,u\rightarrow g\,t)}{dt}\;\; =\;\; 
\frac{g_s^2}{24\,m_t^3}\; \frac{F_{gu}}{u\,t\,s^3\,(s + u)^2\,(t + u)^2}
\; \Gamma (t\,\rightarrow\,u\,g) \;\;\; ,
\label{eq:gu}
\end{equation}
where now we have
\begin{align}
F_{gu} \;=&\;\; 16\,s^8 + s^7\,\left( 241\,t + 72\,u \right)  + 
  4\,s^6\,\left( 305\,t^2 + 222\,t\,u + 34\,u^2 \right)  \nonumber 
\vspace{0.3cm} \\ 
 & + s^5\,\left( 3003\,t^3 + 3750\,t^2\,u + 1367\,t\,u^2 + 136\,u^3 \right) + 
4\,t \,{\left( t + u \right) }^3\, \left( 4\,t^4 + 6\,t^3\,u + 4\,t^2\,u^2 - 
u^4 \right)  \nonumber \vspace{0.3cm} \\ 
& + 2\,s^4\,\left( 1999\,t^4 + 3675\,t^3\,u + 2295\,t^2\,u^2 + 591\,t\,u^3 + 
     34\,u^4 \right)   \nonumber \vspace{0.3cm} \\
& + s^3\,\left( 3003\,t^5 + 7350\,t^4\,u + 6733\,t^3\,u^2 + 3010\,t^2\,u^3 + 
     636\,t\,u^4 + 4\,u^5 \right)  \nonumber \vspace{0.3cm} \\
& + 2\,s^2\,\left( t + u \right) \, \left( 610\,t^5 + 1265\,t^4\,u + 1030\,t^3\,
u^2 + 475\,t^2\,u^3 + 104\,t\,u^4 - 6\,u^5 \right)   \nonumber \vspace{0.3cm} \\
& + s\,{\left( t + u \right) }^2\, \left( 241\,t^5 + 406\,t^4\,u + 314\,t^3\,
u^2 + 148\,t^2\,u^3 + 26\,t\,u^4 - 4\,u^5 \right)  
\end{align}
and once again we obtain a result proportional to $\Gamma (t\,\rightarrow\,u\,
g)$. For the processes $g\,g\,\rightarrow\,t\,\bar{c}$ and $g\,c\,\rightarrow\,
g\,t$ we obtain expressions analogous to~\eqref{eq:gg} and~\eqref{eq:gu}, with 
$\Gamma (t\,\rightarrow\,u\,g)$ replaced by $\Gamma (t\,\rightarrow\,c\,g)$.

If we assume that the branching ratio $BR(t\,\rightarrow\,b\,W)$ is 
approximately 100\% and use $\Gamma (t\,\rightarrow\,b\,W)\,=\,1.42\, \left|
V_{tb}\right|^2$ GeV (a value which includes QCD corrections)~\cite{rev,qcdc}, 
we may express the partial widths of eqs.~\eqref{eq:gg} and~\eqref{eq:gu} as 
$\Gamma (t\,\rightarrow\,q\,g)\,=\,1.42\,\left|V_{tb}\right|^2\,BR(t\, 
\rightarrow\,q\,g)$. In terms of these branching ratios, and using the CTEQ6M 
structure functions~\cite{cteq6}~\footnote{We used a factorization scale equal 
to the mass of the quark top, that being the characteristic scale of these 
reactions. This choice of $\mu_F$ produces smaller cross section values than, 
saying, choosing it equal to the partonic center-of-mass energy~\cite{singt}.} 
to perform the integration in the pdf's, we obtain, for the total cross 
sections, the following results (expressed in picobarn): 
\begin{align}
\sigma(p\,p\,\rightarrow\,g\,g\,\rightarrow\,t\,\bar{q}) &=\;\;\;\left[\,0.5\,
BR (t\,\rightarrow\,u\,g)\;+\; 0.5\,BR(t\,\rightarrow\,c\,g) \right]\,\left|
V_{tb}\right|^2\,10^4 \nonumber \vspace{0.5cm} \\
\sigma(p\,p\,\rightarrow\,g\,g\,\rightarrow\,\bar{t}\,q) &=\;\;\;\sigma(p\,p\,
\rightarrow\,g\,g\,\rightarrow\,t\,\bar{q}) \nonumber \vspace{0.5cm} \\
\sigma(p\,p\,\rightarrow\,g\,q\,\rightarrow\,g\,t) &=\;\;\;\left[\,8.2\,
BR (t\,\rightarrow\,u\,g)\;+\; 0.8\,BR(t\,\rightarrow\,c\,g) \right]\,\left|
V_{tb}\right|^2\,10^4 \nonumber \vspace{0.5cm} \\
\sigma(p\,p\,\rightarrow\,g\,\bar{q}\,\rightarrow\,g\,\bar{t}) &=\;\;\;\left[\,
1.5\,BR(t\,\rightarrow\,u\,g)\;+\; 0.8\,BR(t\,\rightarrow\,c\,g) \right]\,
\left| V_{tb}\right|^2\,10^4 \;\;\; .
\label{eq:sigg}
\end{align} 
These are to be compared with the results obtained in~\cite{nos} for the direct 
top cross section, 
\begin{align}
\sigma(p\,p\,\rightarrow\,g\,q\,\rightarrow\,t) &=\;\;\;\left[\,10.5\,
BR (t\,\rightarrow\,u\,g)\;+\; 1.6\,BR(t\,\rightarrow\,c\,g) \right]\,\left|
V_{tb}\right|^2\,10^4 \nonumber \vspace{0.5cm} \\
\sigma(p\,p\,\rightarrow\,g\,\bar{q}\,\rightarrow\,\bar{t}) &=\;\;\;\left[\,
2.7\, BR(t\,\rightarrow\,u\,g)\;+\; 1.6\,BR(t\,\rightarrow\,c\,g) \right]\,
\left| V_{tb}\right|^2\,10^4 \;\;\; .
\label{eq:sigd}
\end{align}
The larger values of the coefficients affecting the up-quark branching ratios
in eqs.~\eqref{eq:sigg} and~\eqref{eq:sigd} derive from the fact that the pdf 
for that quark is larger than the charm's. The numerical integration has an 
error of less than one percent. Except for the direct top channel, all of these
cross sections (as well as the four-fermion results we will soon present) are
integrated with a cut on the transverse momentum ($p_T$) of the light parton in 
the final state of 15 GeV. This is to remove the collinear and soft 
singularities in the gluon-quark subprocesses to render finite partonic cross
sections, for a finite $p_T$ cut eliminates both of those divergences in 
two-to-two scattering processes. In a realistic analysis including backgrounds, 
a higher $p_T$ cut might well be needed, to suppress background rates in order 
to observe the signal events. That study, however, is beyond the scope of this 
work. Observe how the direct channel cross section is larger than the others. 
Notice, however, that due to the kinematics of that channel, no $p_T$ cut was
applied. When imposing such a cut on the decay products of the top quark 
produced in the direct channel, the corresponding cross section will certainly
be reduced. 

We expect that a cut in $p_T$ should reduce the cross section for the 
gluon-quark channel in a more severe way than the gluon-gluon one. This is due 
to the fact that the $p_T$ cut eliminates most of the soft gluons in the 
gluon-top final state, thus placing us further away from a region where the 
cross section would be larger due to infrared divergencies. In table 1 we show 
the value of the coefficient multiplying
\begin{table}[t]
\begin{center}
\begin{tabular}{cccccc}\hline\hline \\
 Cut in $p_T$ (GeV) & 1 & 5 & 10 & 15 & 20 \\ & & & & & \\ \hline \\
 $g\,u\,\rightarrow\,g\,t$ & 33.4 & 29.3 & 12.0 & 8.2 & 6.4 \vspace{0.2cm} \\
 $g\,g\,\rightarrow\,\bar{u}\,t$ & 1.0 & 0.7 & 0.6 & 0.5 & 0.4 \vspace{0.2cm}
\\\hline\hline\hline
\label{tab:ggq}
\end{tabular}
\caption{Coefficients of $BR (t\,\rightarrow\,u\,g)$ (in picobarn) in 
equations~\eqref{eq:sigg} for several values of the $p_T$ cut.}
\end{center}
\end{table}
the branching ration $BR (t\,\rightarrow\,u\,g)$ in equations~\eqref{eq:sigg},
for the gluon-gluon and gluon-quark channels, for several values of the $p_T$
cut. As expected, the reduction of the gluon-quark cross section is much more
severe than that of the gluon-gluon channel. Nevertheless, a somewhat surprising
feature of these results is the larger values obtained for the channels $g\,q\,
\rightarrow\,g\,t$, compared to the double gluon channel, $g\,g\,\rightarrow\,
\bar{q}\,t$. This runs contrary to the conventional wisdom that the gluon-gluon 
channel ought to be the most important at the LHC, and stems from the fact that 
at the large energy scales expected at the LHC the quark content of the proton 
becomes larger. It also derives from a different factor on the average of the 
initial colours: the double gluon channel requires a colour average factor of
1/64, whereas the gluon-quark channel corresponds to a larger factor of 1/24. 

It is quite remarkable that these cross sections are all proportional to the 
branching ratios for rare decays of the top. These are possible even within the 
SM, at higher orders. For instance, one expects the SM value of $BR(t\,
\rightarrow\,c\,g)$ to be of about $10^{-12}$~\cite{chro,juan}, $BR(t\,
\rightarrow\,u\,g)$ two orders of magnitude smaller. What this means is that, if
whatever new physics lies beyond the SM has no sizeable impact on the flavour 
changing decays of the top quark, so that its branching ratios are not 
substantially different from their SM values, then one does not expect any 
excess of single top production at the LHC through these channels. On the other 
hand, if an excess of single top production is observed, even a small one, the 
expressions~\eqref{eq:sigg} and~\eqref{eq:sigd} tell us that $BR(t\,\rightarrow
\,c\,g)$ and $BR(t\, \rightarrow\,u\,g)$ will have to be very different from 
their SM values. In fact, in models with two Higgs doublets or supersymmetry, 
one expects the branching ratios $BR(t\,\rightarrow\,c\,g)$ and $BR(t\,
\rightarrow\,u\,g)$ to increase immensely~\cite{chro, juan}, in some models 
becoming as large as $\sim 10^{-4}$. If that is the case, eqs.~\eqref{eq:sigg}
and~\eqref{eq:sigd} predict a significant increase in the cross section for 
single top production at the LHC. This cross section is therefore a very 
sensitive observable to probe for new physics. 
 
\section{Four-fermion channels}
\label{sec:4f}

\begin{figure}[ht]
\epsfysize=4cm
\centerline{\epsfbox{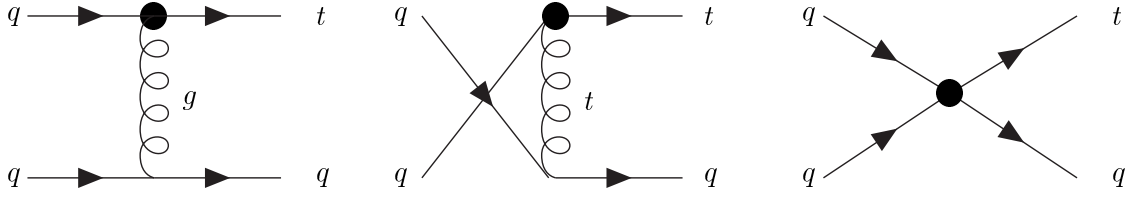}}
\caption{Feynman diagrams for $q\,q\,\rightarrow\,q\,t$. The four-fermion graph
can generate both ``$t$-channel" and ``$u$-channel" contributions.}
\label{fig:qq}
\end{figure}

Other possible processes of single top production involve quark-quark (or 
quark-antiquark) scattering. There are in fact eight such possible reactions, 
which we list in table 2. Notice that in this table we included 
\begin{table}[t]
\begin{center}
\begin{tabular}{cc}\hline\hline \\
 Single top channel & Process number \\ & \\ \hline \\ 
$u\,u\,\rightarrow\,t\,u$  & 1 \vspace{0.2cm} \\ 
$u\,c\,\rightarrow\,t\,c$  & 2 \vspace{0.2cm}\\ 
$u\,\bar{u}\,\rightarrow\,t\,\bar{u}$  & 3 \vspace{0.2cm}\\ 
$u\,\bar{u}\,\rightarrow\,t\,\bar{c}$  & 4 \vspace{0.2cm}\\ 
$u\,\bar{c}\,\rightarrow\,t\,\bar{c}$  & 5 \vspace{0.2cm}\\ 
$d\,\bar{d}\,\rightarrow\,t\,\bar{u}$  & 6 \vspace{0.2cm}\\  
$u\,d\,\rightarrow\,t\,d$  & 7 \vspace{0.2cm}\\ 
$u\,\bar{d}\,\rightarrow\,t\,\bar{d}$  & 8 \\ & 
\\\hline\hline\hline
\label{tab:proc}
\end{tabular}
\caption{List of single top production channels through quark-quark scattering.}
\end{center}
\end{table}
only processes where there is a ``single" flavour violation - in other words,
though processes like $s\,\bar{d}\,\rightarrow\,t\,\bar{u}$ are {\em a priori}
possible from the four-fermion operators we considered, we will not study 
them here. In fact, this is consistent with our choice of gluonic operators, 
as such processes are not possible with the vertices in 
fig.~\eqref{fig:feynrul}. The resulting cross sections now have
contributions from both the gluonic operators~\eqref{eq:lf} and from the four 
fermion operators described earlier. The Feynman diagrams for the process $u\,u
\,\rightarrow\,t\,u$ are shown in figure~\eqref{fig:qq}, but in what concerns
the four-fermion contributions, there is a subtlety that must be mentioned: 
depending on the process considered, each four-fermion operator may contribute
twice to the squared amplitude. For instance, for the process just mentioned, 
there is an operator of Type 1 that surely contributes: $\gamma_{u_1}
\left(\bar{t}\, \lambda^a\,\gamma^{\mu}\, \gamma_R\, u\right)\,\left(\bar{u}
\, \lambda^a\,\gamma_{\mu}\, \gamma_R\, u\right)$. When deducing the Feynman 
rule corresponding to this term, we conclude that it gives us a ``t-channel" 
(when the first $u$ spinor in the operator corresponds to the first incoming 
momentum) and a ``u-channel" (when the first incoming momentum is attributed to 
the second $u$ spinor), both contributions to the amplitude differing by a minus
sign. In fig.~\eqref{fig:qq} we represent only one four-fermion graph for 
simplicity. Notice that for the process $u\,c\,\rightarrow\,t\,c$ the u-channels
we just considered (both from four-fermion operators and gluonic ones) are not
\begin{figure}[ht]
\epsfysize=4cm
\centerline{\epsfbox{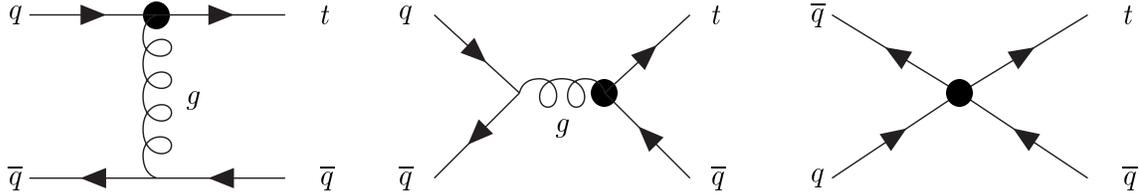}}
\caption{Feynman diagrams for $q\,\bar{q}\,\rightarrow\,\bar{q}\,t$. The 
four-fermion graph can generate $s$, $t$ and $u$ channel contributions.}
\label{fig:qqb}
\end{figure}
present. When antiquarks are present, such as in the process $u\,\bar{u}\,
\rightarrow\,t\,\bar{u}$, gluonic s-channels are also present, as we see in 
fig.~\eqref{fig:qqb}. Again, the four-fermion operators may have several 
distinct contributions to the amplitudes, namely an $s$ channel and a $t$ one. 
One way of thinking of these different amplitude contributions is reading the 
four-fermion operators in terms of interacting currents. Depending on the 
positioning of the fermion spinors within the operator, we will obtain different
currents. For instance, if we number the $u$ spinors in the process just 
mentioned such that $u_1\,\bar{u}_2\,\rightarrow\,t\,\bar{u}_3$ and look at, for
instance, the type 1 four-fermion operators, we see that there are two 
\begin{figure}[ht]
\epsfysize=3cm
\centerline{\epsfbox{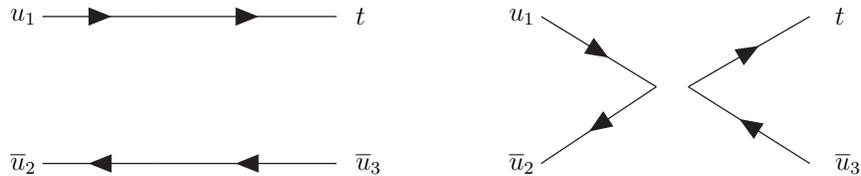}}
\caption{Interpretation of the four-fermion terms contributing to the process
$u_1\,\bar{u}_2\,\rightarrow\,t\,\bar{u}_3$ in terms of currents; notice the 
analog of a $t$-channel and an $s$ one.}
\label{fig:4f}
\end{figure}
possibilities for the disposition of the $u$ spinors, namely $\left(\bar{t}\, 
\lambda^a\,\gamma^{\mu}\, \gamma_R\, u_1\right)\,\left(\bar{u}_2 \, \lambda^a\,
\gamma_{\mu}\, \gamma_R\, u_3\right)$ and $\left(\bar{t}\,\lambda^a\,
\gamma^{\mu}\, \gamma_R\, u_3\right)\,\left(\bar{u}_2 \, \lambda^a\,\gamma_{\mu}
\, \gamma_R\, u_1\right)$, corresponding, in terms of ``currents", to the left
and right diagrams of figure~\eqref{fig:4f}, respectively - a $t$-channel and an
$s$-channel. A careful inspection of the Feynman rules obtained from each of 
these two terms would lead to the conclusion there is a relative minus sign 
between both of their contributions. 

The expressions we obtain are very elaborate, as there are many interference 
terms between the several operators. Let us first consider the terms of the 
squared amplitude that are only proportional to the gluonic couplings. For 
process (1) in table 2 the expression we obtain is
\begin{align} 
|T^{(1)}|^2_{\alpha,\beta} &=  -\,\frac{2}{27\,t\,u}\left\{\,F_1\,\left|
\alpha_{ct} \right|^2\,+\,F_2\,\left|\alpha_{tc}\right|^2\,+\,16\,v^2\,F_3\,
\left(\left|\beta_{ct}\right|^2 + \left|\beta_{tc}\right|^2\right)\,+\,2\,F_4\,
\mbox{Re}(\alpha_{ct}\alpha_{tc}) \right. \nonumber \vspace{1cm} \\
 & \left. \hspace{2cm}\,+\,16\,m_t\,v\,\left[F_5\,\mbox{Im}(\alpha_{ct}
\beta_{tc})\,-\,F_6\,\mbox{Im}(\alpha_{tc}\beta_{tc})\right]\frac{}{} \right\} 
\frac{1}{\Lambda^4}
\label{eq:sig1}
\end{align}
where we have
\begin{align}
F_1\,&=\, 3\, m_t^6\,t - 6\,m_t^4\,t^2 + 6\,m_t^2\,t^3 + 3\,m_t^6\,u + 
12\,m_t^4\,t\,u + 5\, m_t^2\,t^2\,u -  11\,t^3\,u \nonumber \vspace{0.5cm} \\
 & \;\;\;\;- 6\,m_t^4\,u^2 + 5\,m_t^2\,t\,u^2 - 16\,t^2\,u^2 + 6\,m_t^2\,u^3 
- 11\,t\,u^3 \nonumber \vspace{0.5cm} \\
F_2\,&=\,3\,m_t^6\,t - 6\,m_t^4\,t^2 + 6\,m_t^2\,t^3 + 3\,m_t^6\,u - 
20\,m_t^4\,t\,u + 25\,m_t^2\,t^2\,u -  11\,t^3\,u \nonumber \vspace{0.5cm} \\
 & \;\;\;\;- 6\,m_t^4\,u^2 + 25\,m_t^2\,t\,u^2 - 16\,t^2\,u^2 + 
6\,m_t^2\,u^3 - 11\,t\,u^3  \nonumber \vspace{0.5cm} \\
F_3\,&=\,3\,m_t^4\,t - 6\,m_t^2\,t^2 + 6\,t^3 + 3\,m_t^4\,u - 4\,m_t^2\,t\,u 
+ 4\,t^2\,u - 6\,m_t^2\,u^2 + 4\,t\,u^2 + 6\,u^3 \nonumber \vspace{0.5cm} \\
F_4\,&=\,3\,m_t^6\,t - 6\,m_t^4\,t^2 + 6\,m_t^2\,t^3 + 3\,m_t^6\,u - 
4\,m_t^4\,t\,u - 7\,m_t^2\,t^2\,u + 11\,t^3\,u \nonumber \vspace{0.5cm} \\
 & \;\;\;\;- 6\,m_t^4\,u^2 - 7\,m_t^2\,t\,u^2 + 16\,t^2\,u^2 + 
6\,m_t^2\,u^3 + 11\,t\,u^3 \nonumber \vspace{0.5cm} \\
F_5\,&=\,3\,m_t^4\,t - 6\,m_t^2\,t^2 + 6\,t^3 + 3\,m_t^4\,u + 4\,m_t^2\,t\,u - 
t^2\,u - 6\,m_t^2\,u^2 - t\,u^2 + 6\,u^3 \nonumber \vspace{0.5cm} \\
F_6\,&=\,\,3\,\left( t + u \right) \, \left( m_t^4 - 2\,m_t^2\,t + 2\,t^2 -
2\,m_t^2\,u + t\,u + 2\,u^2 \right) \;\;\; .
\end{align}
The expressions we would obtain for process (3) may be obtained from those of 
process (1), with the Mandelstam variable replacements $t\,\rightarrow\,s$ and
$u\,\rightarrow\,t$. For process (2) the expressions are a lot simpler:
\begin{align}
|T^{(2)}|^2_{\alpha,\beta}\; =& -\,\Bigg\{ \frac{2}{9t}\,(s+u)\,(4\,t\,m_t^2 + 
s^2 + u^2)\,\left|\alpha_{ct}\right|^2\,+\,\frac{2}{9t}\,(s + u)\,(s^2 + u^2) 
\,\left|\alpha_{tc}\right|^2 \nonumber \vspace{1cm} \\
 & +\,\frac{32\,v^2}{9t}\,\left[(s + u)\,m_t^2 - 2\,s\,u\right]\,\left(\left|
\beta_{ct}\right|^2 + \left|\beta_{tc}\right|^2\right)\,+\,\frac{4}{9t}\,
(t + m_t^2)\,(s^2 + u^2)\,\mbox{Re}(\alpha_{ct}\alpha_{tc}) \nonumber 
\vspace{1cm} \\
 & +\,\frac{16\,m_t\,v}{9t}\,\left[(m_t^4 - t^2 - 2\,s\,u)\,\mbox{Im}(
\alpha_{ct}\beta_{tc})\,-\,(s^2 + u^2)\,\mbox{Im}(\alpha_{tc} \beta_{tc})\right]
\Bigg\}\,\frac{1}{\Lambda^4} \;\;\; .
\label{eq:exp2}
\end{align}
Processes $\{(5)\,,\,(7)\,,\,(8)\}$ have identical expressions to these for 
process (2); processes $\{(4)\,,\,(6)\}$ have expressions very similar to those 
of eq.~\eqref{eq:exp2}, with the substitution $t\,\leftrightarrow\,s$. 

In table 3, we present the squared amplitude terms involving only 
the 
\begin{table}[t]
\begin{center}
\begin{tabular}{cccc}\hline\hline \\
 Process & $\displaystyle{\frac{|\gamma_{u_1}|^2}{\Lambda^4}}$ & $\displaystyle{
\frac{|\gamma_{u_2}|^2}{\Lambda^4}}$ & $(1 + x^2)\,\displaystyle{\frac{|
\gamma_{u_3}|^2}{\Lambda^4}}$ \\ & & & \\ \hline \\
 (1), (2) & $-\,\displaystyle{\frac{128}{27}}\,s\,(t + u)$ & $\displaystyle{
\frac{8}{9}}\,\left[t^2\,+\,u^2\,-\,m_t^2\,(t + u)\right]$ & 0 \vspace{0.5cm} \\
 (3), (4) &  $-\,\displaystyle{\frac{128}{27}}\,u\,(t + s)$ & 
$\displaystyle{\frac{8}{9}}\,\left[t^2\,+\,s^2\,-\,m_t^2\,(t + s)\right]$ & 0 
\vspace{0.5cm} \\
 (6)  &  $-\,\displaystyle{\frac{256}{27}}\,u\,(t + s)$ &
$\displaystyle{\frac{8}{9}}\,\left[t^2\,+\,s^2\,-\,m_t^2\,(t + s)\right]$ & 0
\vspace{0.5cm} \\
 (6) & $-\,\displaystyle{\frac{32}{9}}\,u\,(s + t)$ & $-\,\displaystyle{
\frac{8}{9}}\,t\,(s + u)$ & $-\,\displaystyle{\frac{16}{27}}\,\left[
8\,t\,s + 3\,u\,(t + s)\right]$ \vspace{0.5cm} \\
 (7) & $-\,\displaystyle{\frac{32}{9}}\,s\,(t + u)$ & $-\,\displaystyle{
\frac{8}{9}}\,u\,(s + t)$ & $-\,\displaystyle{\frac{8}{27}}\,\left[
3\,m_t^2\,(t + u) - 3\,(t^2 + u^2) + 2\,t\,u\right]$ \vspace{0.5cm} \\
 (8) & $-\,\displaystyle{\frac{32}{9}}\,u\,(s + t)$ & $-\,\displaystyle{
\frac{8}{9}}\,s\,(t + u)$ & $-\,\displaystyle{\frac{32}{27}}\,\left[
8\,t\,s + 3\,u\,(t + s)\right]$ \\ & & & 
\\\hline\hline\hline
\label{tab:gaga}
\end{tabular}
\caption{Coefficients of the four-fermion couplings in the squared amplitudes 
for single top quark production. Notice the dependence on $1 + x^2$ on the 
$\gamma_{u_3}$ terms, explained earlier.}
\end{center}
\end{table}
four-fermion couplings. As is obvious from the definitions of the Type 3 
operators (eqs.~\eqref{eq:ga31} and~\eqref{eq:ga32}) they always mix down and up
quarks, thus they have no contribution whatsoever to the processes that involve
only up quarks (processes (1) to (5)). Also, due to the chiral structure of 
the several four-fermion operators, there are no interference terms between 
them. 

To complete the expressions for the squared amplitudes we lack only the 
interference terms between the gluonic operators and the four-fermion ones. We 
present the results for the squared amplitudes in tables 4 and 5. 
Notice the absence of any terms 
\begin{table}[t]
\begin{center}
\begin{tabular}{cccc}\hline\hline \\
 Process & $\displaystyle{\frac{\mbox{Re}(\alpha_{ct}\,\gamma_{u_1})}{
\Lambda^4}}$ &
$\displaystyle{\frac{\mbox{Re}(\alpha_{tc}\,\gamma^*_{u_1})}{\Lambda^4}}$ & 
$\displaystyle{\frac{\mbox{Im}(\beta_{tc}\,\gamma^*_{u_1})}{\Lambda^4}}$ 
\\ & & & \\ \hline \\
 (1) & 0 & 
0 & 
0 \vspace{0.5cm} \\ 
(2) & $-\,\displaystyle{\frac{32}{27}}\,s\,(2\,m_t^2 - s)$ & 
$-\,\displaystyle{\frac{32}{27}}\,s^2$  &
$-\,\displaystyle{\frac{128\,m_t\,v}{27}}\,s$ \vspace{0.5cm} \\
(3) & 0   &
0 &
0 \vspace{0.5cm} \\
 (4) & $-\,\displaystyle{\frac{64}{27}}\,u\,(2\,m_t^2 - u)$ &
$-\,\displaystyle{\frac{64}{27}}\,u^2$    &
$-\,\displaystyle{\frac{256\,m_t\,v}{27}}\,u$   \vspace{0.5cm} \\
 (5) & $-\,\displaystyle{\frac{64}{27}}\,u\,(2\,m_t^2 - u)$ &
$-\,\displaystyle{\frac{64}{27}}\,u^2$    &
$-\,\displaystyle{\frac{256\,m_t\,v}{27}}\,u$   \vspace{0.5cm} \\
 (6) & $-\,\displaystyle{\frac{16}{9}}\,u\,(2\,m_t^2 - u)$ &
$-\,\displaystyle{\frac{16}{9}}\,u^2$    &
$-\,\displaystyle{\frac{64\,m_t\,v}{9}}\,u$   \vspace{0.5cm} \\
 (7) & $-\,\displaystyle{\frac{16}{9}}\,s\,(2\,m_t^2 - s)$ &
$-\,\displaystyle{\frac{16}{9}}\,s^2$    &
$-\,\displaystyle{\frac{64\,m_t\,v}{9}}\,s$   \vspace{0.5cm} \\
 (8) &  $-\,\displaystyle{\frac{16}{9}}\,u\,(2\,m_t^2 - u)$ &
$-\,\displaystyle{\frac{16}{9}}\,u^2$    &
$-\,\displaystyle{\frac{64\,m_t\,v}{9}}\,u$  \\ & & &
\\\hline\hline\hline
\label{tab:gg1}
\end{tabular}
\caption{Interference terms between gluonic and Type 1 four-fermion operators 
for single top production processes.} 
\end{center}
\end{table}
\begin{table}[t]
\begin{center}
\begin{tabular}{cccc}\hline\hline \\
 Process & 
$\displaystyle{\frac{\mbox{Re}(\alpha_{ct}\,\gamma^*_{u_2})}{\Lambda^4}}$ &
$\displaystyle{\frac{\mbox{Re}(\alpha_{tc}\,\gamma^*_{u_2})}{\Lambda^4}}$ & 
$\displaystyle{\frac{\mbox{Im}(\beta_{tc}\,\gamma^*_{u_2})}{\Lambda^4}}$ 
\\ & & & \\ \hline \\
 (1) & $\displaystyle{\frac{8}{27}}\,(m_t^2 + s)\, (t - u)$ &
$\displaystyle{\frac{8}{27}}\,(t^2 - u^2)$ & 
$\displaystyle{\frac{32\,m_t\,v}{27}}\,(t - u)$ \vspace{0.5cm} \\
(2) & $-\,\displaystyle{\frac{8}{27}}\,u\,(2\,m_t^2 - u)$    &
$-\,\displaystyle{\frac{8}{27}}\,u^2$   &
$-\,\displaystyle{\frac{32\,m_t\,v}{27}}\,u$  \vspace{0.5cm} \\
(3) & $\displaystyle{\frac{8}{27}}\,(m_t^2 + u)\, (t - s)$ &
$\displaystyle{\frac{8}{27}}\,(t^2 - s^2)$ &
$\displaystyle{\frac{32\,m_t\,v}{27}}\,(t - s)$ \vspace{0.5cm} \\
 (4) & $\displaystyle{\frac{8}{27}}\,t\,(2\,m_t^2 - t)$    &
$\displaystyle{\frac{8}{27}}\,t^2$    &
$\displaystyle{\frac{32\,m_t\,v}{27}}\,t$   \vspace{0.5cm} \\
 (5) & $\displaystyle{\frac{8}{27}}\,s\,(2\,m_t^2 - s)$    &
$\displaystyle{\frac{8}{27}}\,s^2$    &
$\displaystyle{\frac{32\,m_t\,v}{27}}\,s$   \vspace{0.5cm} \\
 (6) & $\displaystyle{\frac{8}{27}}\,t\,(2\,m_t^2 - t)$    &
$\displaystyle{\frac{8}{27}}\,t^2$    &
$\displaystyle{\frac{32\,m_t\,v}{27}}\,t$   \vspace{0.5cm} \\
 (7) & $\displaystyle{\frac{8}{27}}\,u\,(2\,m_t^2 - u)$    &
$\displaystyle{\frac{8}{27}}\,u^2$    &
$\displaystyle{\frac{32\,m_t\,v}{27}}\,u$   \vspace{0.5cm} \\
 (8) &  $\displaystyle{\frac{8}{27}}\,s\,(2\,m_t^2 - s)$    &
$\displaystyle{\frac{8}{27}}\,s^2$    &
$\displaystyle{\frac{32\,m_t\,v}{27}}\,s$ \\ & & & 
\\\hline\hline\hline
\end{tabular}
\caption{Interference terms between gluonic and Type 2 four-fermion operators 
for single top production processes.} 
\end{center}
\label{tab:glga2}
\end{table}
proportional to $\beta_{ct}$ or $\gamma_{u_3}$, a consequence of the particular 
left-right structures associated with those couplings. The 
equations~\eqref{eq:sig1} and~\eqref{eq:exp2}, and the expressions presented in
tables 3 - 5 refer to the squared amplitudes of the 
several quark-quark processes. Gathering the several multiplicative factors, the
differential cross section is given by
\begin{equation}
\frac{d\,\sigma}{dt}\;\;=\;\; \frac{\alpha_s}{144\,s^2}\,|T|^2\;\;\; ,
\end{equation}
with $|T|^2$ the total squared amplitude for each process and where we have
included a factor of $1/4$ (average on initial spins) and $1/9$ (average on 
initial colours). The overall factor of $\alpha_s$ derives from the fact that 
all squared amplitudes are proportional to $g_s^2$. 

\section{Results for the integrated cross sections}
\label{sec:conc}

We can now gather all the results obtained in this paper and in ref.~\cite{nos}
for the cross sections of single top production. In terms of the couplings, the 
direct channel, eq.~\eqref{eq:sigd}, gives us
\begin{equation}
\sigma_{g\,u\,\rightarrow\,t} \;=\;\left\{ 321\, \left|\alpha_{ut} + 
\alpha^*_{tu}\right|^2\,+\,5080\,\left(\left| \beta_{tu} \right|^2 + \left| 
\beta_{ut} \right|^2\right) + 2556\,\mbox{Im}\left[( \alpha_{ut} + 
\alpha^*_{tu})\, \beta_{tu} \right] \right\}\,\frac{1}{\Lambda^4}\;
\mbox{pb}\;\;\; ,
\end{equation}
for the partonic channel $g\,u\,\rightarrow\,t$. For the gluon-gluon and 
gluon-quark channels, we have, from eqs.~\eqref{eq:sigg},
\begin{align}
\sigma_{g\,g\,\rightarrow\,t\bar{u}} &=\; \left\{\,14\,
\left|\alpha_{ut} + \alpha^*_{tu}\right|^2\,+\,221\,\left(\left|\beta_{tu}
\right|^2 + \left| \beta_{ut} \right|^2\right) + 111\,\mbox{Im}\left[( 
\alpha_{ut} + \alpha^*_{tu})\, \beta_{tu} \right] \right\}\,\frac{1}{\Lambda^4}
\;\mbox{pb} \vspace{0.3cm}\nonumber \\
\sigma_{g\,u\,\rightarrow\,g\,t} &=\; \left\{\,250\,
\left|\alpha_{ut} + \alpha^*_{tu}\right|^2\,+\,3952\,\left(\left|\beta_{tu}
\right|^2 + \left| \beta_{ut} \right|^2\right) + 1988\,\mbox{Im}\left[( 
\alpha_{ut} + \alpha^*_{tu})\, \beta_{tu} \right] \right\}\,
\frac{1}{\Lambda^4}\;\mbox{pb} \;\;\; .
\end{align}

Finally, the four-fermion processes can all be gathered (after integration on
the parton density functions, as before) in a single expression, 
\begin{align}
\sigma^{(u)}_{4F} &=\;\left[\frac{}{}\,171\,\left|\alpha_{ut}
\right|^2\,+\,179\,\left|\alpha_{tu}\right|^2\,-\,176\,\mbox{Re}(\alpha_{ut}\,
\alpha_{tu})\,+\,331\,\mbox{Im}(\alpha_{ut}\,\beta_{tu})\,-\,362\,\mbox{Im}(
\alpha_{tu}\,\beta_{tu}^*)\right. \vspace{0.3cm}\nonumber \\
 & \hspace{0.7cm}+\,689\,\left(\left|\beta_{tu}\right|^2 + \left| 
\beta_{ut} \right|^2\right)\,+\,177\,\mbox{Re}(\alpha_{ut}\,\gamma_{u_1})\,-\,
185\,\mbox{Re}(\alpha_{tu}\,\gamma^*_{u_1})\,-\,16\,\mbox{Im}(\beta_{tu}\,
\gamma^*_{u_1})\vspace{0.6cm}\nonumber \\
 & \hspace{0.7cm}-\,17\,\mbox{Re}(\alpha_{ut}\,\gamma_{u_2})\,+\,17\,
\mbox{Re}(\alpha_{tu}\,\gamma^*_{u_2})\,+\,0.1\,\mbox{Im}(\beta_{tu}\,
\gamma^*_{u_2}) \vspace{0.3cm}\nonumber \\
 & \hspace{0.7cm}+\,\left. 525\,\left|\gamma_{u_1}\right|^2\,+\,94\, \left|
\gamma_{u_2}\right|^2\,+\,88\, \left|\gamma_{u_3}\right|^2 \frac{}{}\right]
\frac{1}{\Lambda^4}\;\mbox{pb} \;\;\; .
\label{eq:sigtu}
\end{align}
For the channels proceeding through the charm quark, we have analogous 
expressions, with different numeric values in most cases due to different parton
content inside the proton. We find
\begin{align}
\sigma_{g\,c\,\rightarrow\,t} &=\;\left\{ \,50\, \left|\alpha_{ct} +
\alpha^*_{tc}\right|^2\,+\,791\,\left(\left| \beta_{tc} \right|^2 + \left|
\beta_{ct} \right|^2\right) + 398\,\mbox{Im}\left[( \alpha_{ct} +
\alpha^*_{tc})\, \beta_{tc} \right] \right\}\,\frac{1}{\Lambda^4}\; \mbox{pb}
\vspace{2cm} \nonumber \\
\sigma_{g\,g\,\rightarrow\,t\bar{c}} &=\; \left\{\,14\,
\left|\alpha_{ct} + \alpha^*_{tc}\right|^2\,+\,221\,\left(\left|\beta_{tc}
\right|^2 + \left| \beta_{ct} \right|^2\right) + 111\,\mbox{Im}\left[(
\alpha_{ct} + \alpha^*_{tc})\, \beta_{tc} \right] \right\}\,\frac{1}{\Lambda^4}
\;\mbox{pb} \vspace{1cm}\nonumber \\
\sigma_{g\,c\,\rightarrow\,g\,t} &=\; \left\{\,25\,
\left|\alpha_{ct} + \alpha^*_{tc}\right|^2\,+\,395\,\left(\left|\beta_{tc}
\right|^2 + \left| \beta_{ct} \right|^2\right) + 199\,\mbox{Im}\left[(
\alpha_{ct} + \alpha^*_{tc})\, \beta_{tc} \right] \right\}\, \frac{1}{\Lambda^4}
\;\mbox{pb} \vspace{1cm} \nonumber \\
\sigma^{(c)}_{4F} &=\;\left[\frac{}{}\,20\,\left|\alpha_{ct}
\right|^2\,+\,20\,\left|\alpha_{tc}\right|^2\,-\,12\,\mbox{Re}(\alpha_{ct}\,
\alpha_{tc})\,+\,55\,\mbox{Im}(\alpha_{ct}\,\beta_{tc})\,-\,53\,\mbox{Im}(
\alpha_{tc}\,\beta_{tc}^*)\right. \vspace{0.3cm}\nonumber \\
 & \hspace{0.7cm}+\,107\,\left(\left|\beta_{tc}\right|^2 + \left|
\beta_{ct} \right|^2\right)\,+\,41\,\mbox{Re}(\alpha_{ct}\,\gamma_{c_1})\,-\,
41\,\mbox{Re}(\alpha_{tc}\,\gamma^*_{c_1})\,+\,0.2\,\mbox{Im}(\beta_{tc}\,
\gamma^*_{c_1})\vspace{0.6cm}\nonumber \\
 & \hspace{0.7cm}-\,3\,\mbox{Re}(\alpha_{ct}\,\gamma_{c_2})\,+\,3\,
\mbox{Re}(\alpha_{tc}\,\gamma^*_{c_2})\,-\,0.5\,\mbox{Im}(\beta_{tc}\,
\gamma^*_{c_2}) \vspace{0.3cm}\nonumber \\
 & \hspace{0.7cm}+\,\left. 95\,\left|\gamma_{c_1}\right|^2\,+\,24\, \left|
\gamma_{c_2}\right|^2\,+\,27\, \left|\gamma_{c_3}\right|^2 \frac{}{}\right]
\frac{1}{\Lambda^4}\;\mbox{pb} \;\;\; .
\label{eq:sigtc}
\end{align}
Within the four-fermion cross sections, eqs.~\eqref{eq:sigtu} 
and~\eqref{eq:sigtc}, are the results for production of a bottom quark 
alongside the top, through the processes $u\,b\,\rightarrow\,t\,b$ and $u\,
\bar{b}\,\rightarrow\,t\,\bar{b}$ (and analogous processes for the $c$ quark). 
They are given by
\begin{align}
\sigma^{(u)}_{t + b} &=\;\left[\frac{}{}\,8\,\left|\alpha_{ut}
\right|^2\,+\,9\,\left|\alpha_{tu}\right|^2\,-\,2\,\mbox{Re}(\alpha_{ut}\,
\alpha_{tu})\,+\,28\,\mbox{Im}(\alpha_{ut}\,\beta_{tu})\,-\,32\,\mbox{Im}(
\alpha_{tu}\,\beta_{tu}^*)\right. \vspace{0.3cm}\nonumber \\
 & \hspace{0.7cm}+\,59\,\left(\left|\beta_{tu}\right|^2 + \left|
\beta_{ut} \right|^2\right)\,+\,12\,\mbox{Re}(\alpha_{ut}\,\gamma_{u_1})\,-\,
13\,\mbox{Re}(\alpha_{tu}\,\gamma^*_{u_1})\,-\,3\,\mbox{Im}(\beta_{tu}\,
\gamma^*_{u_1})\vspace{0.6cm}\nonumber \\
 & \hspace{0.7cm}-\,2\,\mbox{Re}(\alpha_{ut}\,\gamma_{u_2})\,+\,2\,
\mbox{Re}(\alpha_{tu}\,\gamma^*_{u_2})\,+\,0.5\,\mbox{Im}(\beta_{tu}\,
\gamma^*_{u_2}) \vspace{0.3cm}\nonumber \\
 & \hspace{0.7cm}+\,\left. 19\,\left|\gamma_{u_1}\right|^2\,+\,5\, \left|
\gamma_{u_2}\right|^2\,+\,16\, \left|\gamma_{u_3}\right|^2 \frac{}{}\right]
\frac{1}{\Lambda^4}\;\mbox{pb} 
\end{align}
and
\begin{align}
\sigma^{(c)}_{t + b} &=\;\left[\frac{}{}\,0.4\,\left|\alpha_{ct}
\right|^2\,+\,0.6\,\left|\alpha_{tc}\right|^2\,+\,0.2\,\mbox{Re}(\alpha_{ct}\,
\alpha_{tc})\,+\,2\,\mbox{Im}(\alpha_{ct}\,\beta_{tc})\,-\,3\,\mbox{Im}(
\alpha_{tc}\,\beta_{tc}^*)\right. \vspace{0.3cm}\nonumber \\
 & \hspace{0.7cm}\left. +\,5\,\left(\left|\beta_{tc}\right|^2 + \left|
\beta_{ct} \right|^2\right)\,+\,\left|\gamma_{c_1}\right|^2\,+\,0.2\, 
\left| \gamma_{c_2}\right|^2\,+\,0.6\, \left|\gamma_{c_3}\right|^2 \frac{}{}
\right] \frac{1}{\Lambda^4}\;\mbox{pb} 
\end{align}
where the interference terms between the $\{\alpha\,,\,\beta\}$ and the $\gamma$
were left out because they were too small when compared with the remaining 
terms.

Finally, by changing the pdf integrations and using the second vertex in 
figure~\eqref{fig:feynrul}, we can also obtain the cross sections for anti-top 
production. We obtain
\begin{align}
\sigma_{g\,\bar{u}\,\rightarrow\,\bar{t}} &=\;\left\{\,83\,\left|\alpha_{ut} +
\alpha^*_{tu}\right|^2\,+\,1312\,\left(\left| \beta_{tu} \right|^2 + \left|
\beta_{ut} \right|^2\right) + 660\,\mbox{Im}\left[( \alpha_{ut} +
\alpha^*_{tu})\, \beta_{tu} \right] \right\}\,\frac{1}{\Lambda^4}\; \mbox{pb}
\vspace{2cm} \nonumber \\
\sigma_{g\,g\,\rightarrow\,\bar{t}\,u} &=\; \left\{\,14\,
\left|\alpha_{ut} + \alpha^*_{tu}\right|^2\,+\,221\,\left(\left|\beta_{tu}
\right|^2 + \left| \beta_{ut} \right|^2\right) + 111\,\mbox{Im}\left[(
\alpha_{ut} + \alpha^*_{tu})\, \beta_{tu} \right] \right\}\,\frac{1}{\Lambda^4}
\;\mbox{pb} \vspace{1cm}\nonumber \\
\sigma_{g\,\bar{u}\,\rightarrow\,g\,\bar{t}} &=\; \left\{\,45\,
\left|\alpha_{ut} + \alpha^*_{tu}\right|^2\,+\,711\,\left(\left|\beta_{tu}
\right|^2 + \left| \beta_{ut} \right|^2\right) + 358\,\mbox{Im}\left[(
\alpha_{ut} + \alpha^*_{tu})\, \beta_{tu} \right] \right\}\, \frac{1}{\Lambda^4}
\;\mbox{pb} \vspace{1cm} \nonumber \\
\sigma^{(u)}_{4F} &=\;\left[\frac{}{}\,32\,\left|\alpha_{ut}
\right|^2\,+\,32\,\left|\alpha_{tu}\right|^2\,-\,19\,\mbox{Re}(\alpha_{ut}\,
\alpha_{tu})\,+\,90\,\mbox{Im}(\alpha_{ut}\,\beta_{tu})\,-\,90\,\mbox{Im}(
\alpha_{tu}\,\beta_{tu}^*)\right. \vspace{0.3cm}\nonumber \\
 & \hspace{0.7cm}+\,178\,\left(\left|\beta_{tu}\right|^2 + \left|
\beta_{ut} \right|^2\right)\,-\,21\,\mbox{Re}(\alpha_{ut}\,\gamma_{u_1})\,+\,
21\,\mbox{Re}(\alpha_{tu}\,\gamma^*_{u_1})\,+\,\mbox{Im}(\beta_{tu}\,
\gamma^*_{u_1})\vspace{0.6cm}\nonumber \\
 & \hspace{0.7cm}+\,3\,\mbox{Re}(\alpha_{ut}\,\gamma_{u_2})\,-\,1\,
\mbox{Re}(\alpha_{tu}\,\gamma^*_{u_2})\,-\,\mbox{Im}(\beta_{tu}\,
\gamma^*_{u_2}) \vspace{0.3cm}\nonumber \\
 & \hspace{0.7cm}+\,\left. 56\,\left|\gamma_{u_1}\right|^2\,+\,26\, \left|
\gamma_{u_2}\right|^2\,+\,35\, \left|\gamma_{u_3}\right|^2 \frac{}{}\right]
\frac{1}{\Lambda^4}\;\mbox{pb} 
\label{eq:sigatu}
\end{align}
and also
\begin{align}
\sigma_{g\,\bar{c}\,\rightarrow\,\bar{t}} &=\;\left\{\,50\,\left|\alpha_{ct} +
\alpha^*_{tc}\right|^2\,+\,791\,\left(\left| \beta_{tc} \right|^2 + \left|
\beta_{ct} \right|^2\right) + 398\,\mbox{Im}\left[( \alpha_{ct} +
\alpha^*_{tc})\, \beta_{tc} \right] \right\}\,\frac{1}{\Lambda^4}\; \mbox{pb}
\vspace{2cm} \nonumber \\
\sigma_{g\,g\,\rightarrow\,\bar{t}\,c} &=\; \left\{\,14\,
\left|\alpha_{ct} + \alpha^*_{tc}\right|^2\,+\,221\,\left(\left|\beta_{tc}
\right|^2 + \left| \beta_{ct} \right|^2\right) + 111\,\mbox{Im}\left[(
\alpha_{ct} + \alpha^*_{tc})\, \beta_{tc} \right] \right\}\,\frac{1}{\Lambda^4}
\;\mbox{pb} \vspace{1cm}\nonumber \\
\sigma_{g\,\bar{c}\,\rightarrow\,g\,\bar{t}} &=\; \left\{\,25\,
\left|\alpha_{ct} + \alpha^*_{tc}\right|^2\,+\,395\,\left(\left|\beta_{tc}
\right|^2 + \left| \beta_{ct} \right|^2\right) + 199\,\mbox{Im}\left[(
\alpha_{ct} + \alpha^*_{tc})\, \beta_{tc} \right] \right\}\, \frac{1}{\Lambda^4}
\;\mbox{pb} \vspace{1cm} \nonumber \\
\sigma^{(c)}_{4F} &=\;\left[\frac{}{}\,20\,\left|\alpha_{ct}
\right|^2\,+\,20\,\left|\alpha_{tc}\right|^2\,-\,12\,\mbox{Re}(\alpha_{ct}\,
\alpha_{tc})\,+\,53\,\mbox{Im}(\alpha_{ct}\,\beta_{tc})\,-\,55\,\mbox{Im}(
\alpha_{tc}\,\beta_{tc}^*)\right. \vspace{0.3cm}\nonumber \\
 & \hspace{0.6cm}+\,107\,\left(\left|\beta_{tc}\right|^2 + \left|
\beta_{ct} \right|^2\right)\,-\,32\,\mbox{Re}(\alpha_{ct}\,\gamma_{c_1})\,+\,
36\,\mbox{Re}(\alpha_{tc}\,\gamma^*_{c_1})\,+\,8\,\mbox{Im}(\beta_{tc}\,
\gamma^*_{c_1})\vspace{0.6cm}\nonumber \\
 & \hspace{0.6cm}+\,7\,\mbox{Re}(\alpha_{ct}\,\gamma_{c_2})\,-\,7\,
\mbox{Re}(\alpha_{tc}\,\gamma^*_{c_2}) \,+\,0.3\,\mbox{Im}(\beta_{tc}\,
\gamma^*_{c_2})\vspace{0.6cm}\nonumber \\
 & \hspace{0.6cm}+\,82\,\left|\gamma_{c_1}\right|^2\,+\,
29\, \left| \gamma_{c_2}\right|^2\left.\,+\,29\, \left|\gamma_{c_3}\right|^2 
\frac{}{}\right] \frac{1}{\Lambda^4}\;\mbox{pb} \;\;\; .
\label{eq:sigatc}
\end{align}
For completeness, the cross sections for production of an anti-top alongside 
with a bottom quark are (leaving out terms which are too small compared with 
the others)
\begin{align}
\sigma^{(u)}_{\bar{t} + b} &=\;\left[\frac{}{}\,1\,\left|\alpha_{ut}
\right|^2\,+\,1\,\left|\alpha_{tu}\right|^2\,+\,0.4\,\mbox{Re}(\alpha_{ut}\,
\alpha_{tu})\,+\,5\,\mbox{Im}(\alpha_{ut}\,\beta_{tu})\,-\,4\,\mbox{Im}(
\alpha_{tu}\,\beta_{tu}^*)\right. \vspace{0.3cm}\nonumber \\
 & \hspace{0.7cm}+\,10\,\left(\left|\beta_{tu}\right|^2 + \left|
\beta_{ut} \right|^2\right)\,-\,2\,\mbox{Re}(\alpha_{ut}\,\gamma_{u_1})\,+\,
1\,\mbox{Re}(\alpha_{tu}\,\gamma^*_{u_1})\,-\,0.5\,\mbox{Im}(\beta_{tu}\,
\gamma^*_{u_1})\vspace{0.6cm}\nonumber \\
 & \hspace{0.7cm} +\,\,0.2\,\mbox{Re}(\alpha_{ut}\,\gamma^*_{u_2})\,-\,0.2
\,\mbox{Re}(\alpha_{tu}\,\gamma^*_{u_2})\,+\,2\, \left| \gamma_{u_1}\right|^2
\vspace{0.6cm}\nonumber \\
 & \hspace{0.7cm}\left.+\,0.5\, \left|\gamma_{u_2}\right|^2 \,+\,2\, 
\left|\gamma_{u_3}\right|^2 \frac{}{}\right]
\frac{1}{\Lambda^4}\;\mbox{pb} \vspace{0.3cm}\nonumber \\
\sigma^{(c)}_{\bar{t} + b} &=\;\left[\frac{}{}\,0.4\,\left|\alpha_{ct}
\right|^2\,+\,0.6\,\left|\alpha_{tc}\right|^2\,+\,0.2\,\mbox{Re}(\alpha_{ct}\,
\alpha_{tc})\,+\,3\,\mbox{Im}(\alpha_{ct}\,\beta_{tc})\,-\,2\,\mbox{Im}(
\alpha_{tc}\,\beta_{tc}^*)\right. \vspace{0.3cm}\nonumber \\
 & \hspace{0.7cm} +\,5\,\left(\left|\beta_{tc}\right|^2 + \left|
\beta_{ct} \right|^2\right)\,-\,0.7\,\,\mbox{Re}(\alpha_{ct}\,\gamma_{c_1})\,+\,
0.5\,\mbox{Re}(\alpha_{tc}\,\gamma^*_{c_1})\,-\,0.5\,\mbox{Im}(\beta_{tc}\,
\gamma^*_{c_1})\vspace{0.6cm}\nonumber \\
 & \hspace{0.7cm} +\,\left. 0.8\,\left|\gamma_{c_1}\right|^2\,+\,0.2\, \left|
\gamma_{c_2}\right|^2\,+\,0.6\, \left|\gamma_{c_3}\right|^2 \frac{}{}\right]
\frac{1}{\Lambda^4}\;\mbox{pb} \;\;\; .
\end{align} 

We have thus far presented the complete expressions for the cross sections but,
as was discussed earlier and is made manifest by equation~\eqref{eq:rel}, some
of the operators we considered are not independent. In fact, eq.~\eqref{eq:rel}
implies that we can choose two of the couplings $\{\alpha_{ut}\,,\,\alpha_{tu}\,
,\,\beta_{ut}\,,\,\beta_{tu}\,,\,\gamma_{u_1}\}$ to be equal to zero. Notice
that $\gamma_{u_2}$ and $\gamma_{u_3}$ are not included in this choice, as the
respective operators do not enter into equations~\eqref{eq:rel}. A similar
conclusion may be drawn, of course, about the couplings $\{\alpha_{ct}\,,\,
\alpha_{tc}\,,\,\beta_{ct}\,,\,\beta_{tc}\,,\,\gamma_{c_1}\}$. We choose to 
set $\beta_{tu}$ and $\gamma_{u_1}$ to zero, as this choice eliminates many of
the interference terms of the cross sections. Summing all of the different 
contributions, we obtain, for the single top production cross section, the 
following results:
\begin{align}
\sigma^{(u)}_{single\;\, t} &=\;\left[\frac{}{}\,756\,\left|\alpha_{ut}
\right|^2\,+\,764\,\left|\alpha_{tu}\right|^2\,+\,994\,\mbox{Re}(\alpha_{ut}\,
\alpha_{tu})\,+\,9942\,\left|\beta_{ut}\right|^2 \right. \vspace{0.3cm}
\nonumber \\
 & \hspace{0.7cm}\left. -\,17\,\mbox{Re}(\alpha_{ut}\,\gamma_{u_2})\,+\,17\,
\mbox{Re}(\alpha_{tu}\,\gamma^*_{u_2})\,+\,94\, \left|
\gamma_{u_2}\right|^2\,+\,88\, \left|\gamma_{u_3}\right|^2 \frac{}{}\right]
\frac{1}{\Lambda^4}\;\mbox{pb} \;\;\; ,
\nonumber \vspace{0.3cm} \\
\sigma^{(c)}_{single\;\, t} &=\;\left[\frac{}{}\,109\,\left|\alpha_{ct}
\right|^2\,+\,109\,\left|\alpha_{tc}\right|^2\,+\,166\,\mbox{Re}(\alpha_{ct}\,
\alpha_{tc})\,+\,1514\,\left|\beta_{ct}\right|^2 \right. \vspace{0.3cm}
\nonumber \\
 & \hspace{0.7cm}\left. -\,3\,\mbox{Re}(\alpha_{ct}\,\gamma_{c_2})\,+\,3\,
\mbox{Re}(\alpha_{tc}\,\gamma^*_{c_2})\,+\,24\, \left|
\gamma_{c_2}\right|^2\,+\,27\, \left|\gamma_{c_3}\right|^2 \frac{}{}\right]
\frac{1}{\Lambda^4}\;\mbox{pb} \;\;\; .
\label{eq:res}
\end{align}
For anti-top production,
\begin{align}
\sigma^{(u)}_{single\;\, \bar{t}} &=\;\left[\frac{}{}\,174\,\left|\alpha_{ut}
\right|^2\,+\,174\,\left|\alpha_{tu}\right|^2\,+\,265\,\mbox{Re}(\alpha_{ut}\,
\alpha_{tu})\,+\,2422\,\left|\beta_{ut}\right|^2 \right. \vspace{0.3cm}
\nonumber \\
 & \hspace{0.7cm}\left. +\,3\,\mbox{Re}(\alpha_{ut}\,\gamma_{u_2})\,-\,
\mbox{Re}(\alpha_{tu}\,\gamma^*_{u_2})\,+\,26\, \left|
\gamma_{u_2}\right|^2\,+\,35\, \left|\gamma_{u_3}\right|^2 \frac{}{}\right]
\frac{1}{\Lambda^4}\;\mbox{pb} \;\;\; , \nonumber \vspace{0.3cm} \\
\sigma^{(c)}_{single\;\, \bar{t}} &=\left[\frac{}{}\,109\,\left|\alpha_{ct}
\right|^2\,+\,109\,\left|\alpha_{tc}\right|^2\,+\,166\,\mbox{Re}(\alpha_{ct}\,
\alpha_{tc})\,+\,1514\,\left|\beta_{ct}\right|^2 \right. \vspace{0.3cm}
\nonumber \\
 & \hspace{0.7cm}\left. +\,7\,\mbox{Re}(\alpha_{ct}\,\gamma_{c_2})\,-\,7\,
\mbox{Re}(\alpha_{tc}\,\gamma^*_{c_2})\,+\,29\, \left|
\gamma_{c_2}\right|^2\,+\,29\, \left|\gamma_{c_3}\right|^2 \frac{}{}\right]
\frac{1}{\Lambda^4}\;\mbox{pb} \;\;\; .
\end{align}

\begin{figure}[ht]
\epsfysize=10cm
\centerline{\epsfbox{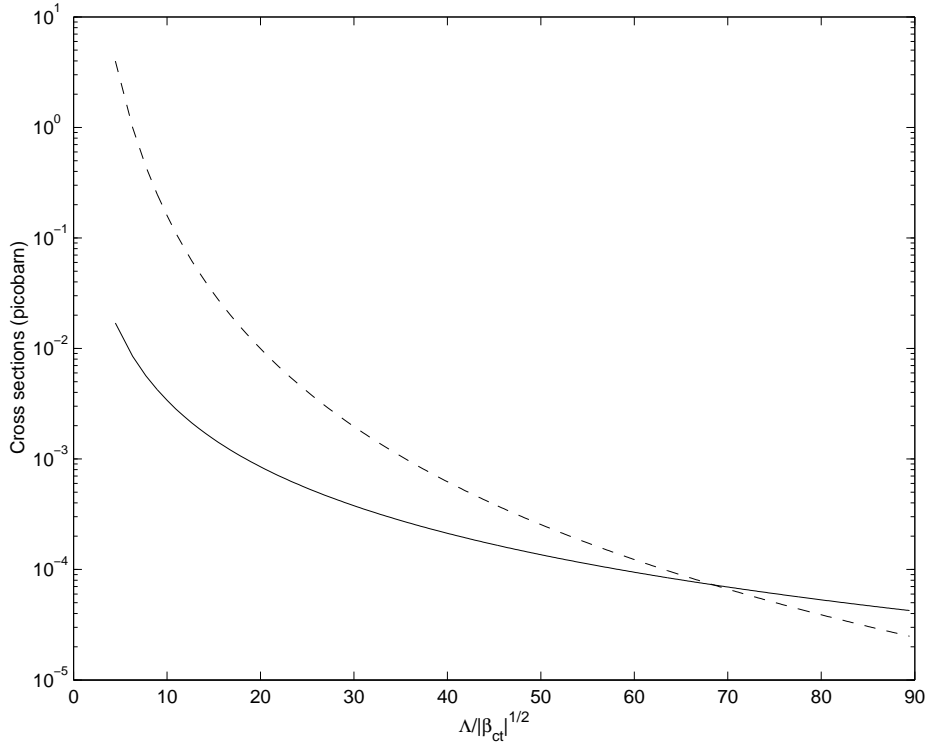}}
\caption{Interference cross section (solid line) and total cross section, 
from eq.~\eqref{eq:res} (dashed line), versus $\Lambda/\sqrt{|\beta_{ct}|}$.
Notice that the two curves only cross for very large values of the new physics 
energy scale.}
\label{fig:fig}
\end{figure}

As mentioned earlier, there are also interference terms between our gluonic
operators and the electroweak processes of single top production in the SM. 
These depend on $\Lambda^{-2}$ but are very small, namely
\begin{equation}
\sigma^{int}_{single\;\, t} \;=\; 0.81\,\frac{|\mbox{Re}(
\beta_{ut})|}{\Lambda^2} \,+\,0.27\,\frac{|\mbox{Re}(\beta_{ct})|}{\Lambda^2}
\;\;\; \mbox{pb}. 
\label{eq:ints}
\end{equation}
Given the different $\Lambda$ dependence on these interference terms and 
our results for the cross sections~\eqref{eq:res}, it is worth asking what is 
the domain of values of $\Lambda$ for which the $\Lambda^{-4}$ terms are 
superior to those of eq.~\eqref{eq:ints}. To obtain a reasonable estimate, we 
consider only the terms proportional to $|\beta|^2$ in the cross 
sections~\eqref{eq:res} and~\eqref{eq:ints}. We further simplify the estimation by taking $\beta_{ut}\,=\,\beta_{ct}\,|V_{ub}/V_{cb}|$. In 
figure~\eqref{fig:fig} we plot the value of the cross  
sections~\eqref{eq:res} and~\eqref{eq:ints} versus 
$\Lambda/\sqrt{|\beta_{ct}|}$. We see that they only cross for very large values
of this variable, of about 70 TeV. For scales of new physics inferior to, say, 
$\sim$ 30 TeV, the cross sections presented in this paper are almost
exactly the full contributions from the set of effective operators we chose to
single top production.  

There is an extensive literature on the subject of single top 
production~\cite{topcr}. For the LHC, the SM prediction is usually considered to
be $319.7\,\pm\,19.3$ pb~\cite{singt}. Considering the large numbers we are 
obtaining in the expressions above - specially the coefficients of the $\beta$
couplings, though the others are not in any way negligible - we can see that 
even a small deviation from the SM framework will produce a potentially large
effect in this cross section. It is indeed a good observable to test new 
physics, as it seems so sensible to its presence. Alternatively, if the cross 
section for single top production at the LHC is measured in the years to come 
and is found to be in complete agreement with the SM predicted value, then we 
will be able to set extremely stringent bounds on the couplings $\{\alpha\,,\,
\beta\,,\,\gamma \}$ - on new physics in general - precisely for the same 
reasons. 

In conclusion, we have calculated the contributions from a large set of 
dimension six operators to cross sections of several processes of single top 
production at the LHC. All cross sections involving gluons in the initial or 
final states are proportional to branching ratios of rare top quark decays. This
makes these processes extremely sensitive to new physics, since those branching 
ratios may vary by as much as eight orders of magnitude in the SM and extended 
models. The four-fermion operators we chose break this proportionality so that,
even if the branching ratios of the top quark conform to those of the SM, we may
still have an excess of single top production at the LHC, stemming from those 
same operators. One of the advantages of working in a fully gauge-invariant 
manner is the possibility of using the equations of motion to introduce 
relations between the operators and thus reduce the number of independent 
parameters. One possible further simplification, if one so wishes, would be to
consider each generation's couplings related by the SM CKM matrix elements, so
that, for instance, $\alpha_{tu}\,=\,\alpha_{tc}\,|V_{ub}/V_{cb}|$. This should 
constitute a reasonable estimate of the difference in magnitude between each
generations' couplings. Finally, in this paper we presented both the total 
anomalous cross sections for single top production and those of the individual
processes that contribute to it. If there is any experimental method - through
kinematical cuts or jet analysis - to distinguish between each of the possible 
partonic channels (direct top production; gluon-quark fusion; gluon-gluon 
fusion; quark-quark scattering), the several expressions we presented here will
allow a direct comparison between theory and experiment. At this point a 
thorough detector simulation of these processes is needed to establish under 
which conditions, if any, they might be observed at the LHC, and what precision 
one might expect to obtain on bounds on the couplings $\{\alpha\,,\,\beta\,,\, 
\gamma\}$. 

\vspace{0.25cm}
{\bf Acknowledgments:} Our thanks to Augusto Barroso, C.-P. Yuan and our
colleagues from LIP for valuable discussions. Special thanks to O. Oliveira for 
early contributions to this work. Our further thanks to A.B. and 
Ant\'onio Onofre for a careful reading of the manuscript. This work is 
supported by Funda\c{c}\~ao para a Ci\^encia e Tecnologia under contract 
PDCT/FP/FNU/50155/2003 and POCI/FIS/59741/2004. P.M.F. is supported by FCT 
under contract SFRH/BPD/5575/2001.

\end{document}